\documentclass[11pt,palatino]{article}
\usepackage[numbers,sort&compress,square]{natbib}
\usepackage{amsfonts,amsmath,amssymb,bm,url}
\usepackage{graphicx,epsfig,subfigure}
\begin{document}

\title{%
The 
Difficulties %Challenges 
of Addressing Interdisciplinary Challenges at the Foundations of Data Science%
\footnote{Appearing in SIAM News, SIGACT News, etc.}
}
\author{
Michael W. Mahoney\thanks{ICSI and Department of Statistics,
                          University of California Berkeley,
                          Berkeley, CA 94720.
                          Email: \texttt{mmahoney@stat.berkeley.edu} }
}
\date{}
\maketitle

% %\vspace{-5mm}
% \begin{center}
% \emph{
% \textbf{This is a draft: from \today. \\ Do not distribute.}
% }
% \end{center}
% %\vspace{0.1in}

\noindent
The Transdisciplinary Research in Principles of Data Science (TRIPODS) program is an effort funded by the National Science Foundation (NSF) to establish the foundations of data science.
It aims to unite the statistics, mathematics, and theoretical computer science research communities (three areas central to the foundations of data, hence the ``TRI'' in the name) to build the theoretical foundations that will enable continued data-driven discovery and breakthroughs in the natural and social sciences, engineering, and beyond.  

The size and scope of its interdisciplinarity make TRIPODS an unusual program. 
The program's first phase consists of \$17.7 million in funding for 12 Phase~I institutes.\footnote{\url{https://nsf-tripods.org/institutes/}}  
After an initial three-year effort (currently in progress), a longer second phase will consist of a smaller number of larger full-scale Phase~II institutes.
Additionally, a related TRIPODS+X program is designed to expand the scope of the original TRIPODS Phase~I projects to involve interactions with  researchers from domain ``X,'' where ``X'' is astronomy, genetics, materials science, neuroscience, the social sciences, or any one of a wide range of other data-driven disciplines.
Spurred by the success of TRIPODS and the excitement of using it as a model to fund interdisciplinary research more generally, the NSF also recently announced the creation of a second parallel TRIPODS Phase~I--Phase~II program, with this new cohort also including electrical engineering.

I am the director and principal investigator (PI) of the new University of California at Berkeley Foundations of Data Analysis (FODA) Institute,\footnote{\url{https://foda.berkeley.edu/}} one of the 12 original TRIPODS Phase~I institutes.
I work with with my co-PIs---Peter Bartlett, Michael Jordan, Fernando Perez, Bin Yu, and Uros Seljak (with TRIPODS+X)---to make transformational advances in the interdisciplinary foundations of data science, including both teaching and research.
We work with a range of campus partners---including the Berkeley Institute for Data Science (BIDS), Simons Institute for the Theory of Computing, Real-time Intelligent Secure Explainable (RISE) Lab, and Lawrence Berkeley National Laboratory (LBNL)---who address other complementary aspects of data science.
Beyond simply proving theorems, we are interested in how the foundations interact synergistically with increasingly data-driven domain sciences.
For me, these efforts follow a long line of previous work, including the Workshop on Algorithms for Modern Massive Data Set (MMDS) meetings~\cite{MMDS06,MMDS08_KDDExp} and the 2016 Park City Mathematics Institute summer school on ``The Mathematics of Data''~\cite{PCMI_math_of_data_BOOK}.

The TRIPODS program is both timely and important.
Every field needs its foundations to understand \emph{when} and \emph{why} methods work as they do, but the three areas that TRIPODS identifies  as being closest to the foundations of data science have very different foundational principles.
Much data science education and training is currently limited to teaching tools (Python routines, etc.), rather than an understanding of foundational principles.
While foundational research may result in plainer graphics and be less immediately applicable than work in other more applied areas, failing to invest in foundational questions will lead to fields that are less intellectually rich, with a hollower shell.  In such scenarios, deeper connections between superficially-different methods applied in very different areas will not be recognized, understood, or exploited.

Furthermore, the TRIPODS program is timely and important since it itself provides a model for funding cross-cutting research more generally.
The interdisciplinary challenges in orchestrating fruitful interactions between foundational computer scientists, statisticians, and applied mathematicians will be mirrored---and probably much more so---when considering social, behavioral, and economic challenges associated with large-scale computing platforms; ethical and responsible uses of data; machine learning for materials science or the biomedical sciences; or any more outward-facing challenges important to data science.

In this article, I will discuss experiences with several of these challenges.

%\vspace{-3mm}
\paragraph{Community Background.}

Beyond facing difficult technical questions about the foundations of data, building cross-cutting platforms between different disciplines and conducting 
truly   
interdisciplinary research is \emph{very} challenging.
Funding mechanisms, hiring mechanisms, and differing disciplinary cultures all conspire against this.
If the goal is to bridge the gap between different research communities, then understanding the communities' backgrounds helps us identify three key classes of challenges that arise in such efforts.

\textbf{Structural Challenge.}
TRIPODS stems in part from a 
phenomenon,  
a sort of ``structural chasm,'' that can cause interdisciplinary work to ``fall between the cracks.''
Scientists conducting research that cuts across traditional community disciplines are familiar with the effects of this occurrence.
If one has a proposal that sits squarely between NSF's Division of Mathematical Sciences (DMS, which funds mathematics and statistics research) and its Directorate for Computer and Information Science and Engineering (CISE, which funds computer science research), then he/she must decide where to submit the proposal.
Upon sending it to DMS, reviewers said ``Great ideas, and high impact, but this isn't quite within the scope of DMS---send it to CISE.''
After sending that same proposal to CISE, reviewers said ``Great ideas, and high impact, but this isn't quite within the scope of CISE---send it to DMS.'' 
Relatedly, although universities are great at putting together interdisciplinary teams, they are much less adept at hiring interdisciplinary people.
This sort of \emph{structural challenge}, whereby newly-forming areas do not conform well to existing administrative silos, is perhaps the most obvious type of issue  to arise in interdisciplinary~efforts.

\textbf{Justification Challenge.}
Motivated by this predicament, 
and prompted by the NSF,
Petros Drineas (Purdue University) and Xiaoming Huo (Georgia Tech) organized an exploratory workshop on ``Theoretical Foundations of Data Science (TFoDS): Algorithmic, Mathematical, and Statistical''\footnote{\url{http://www.cs.rpi.edu/TFoDS}} in April 2016. 
The workshop's objective was ``to identify important research challenges that strengthen and broaden the mathematical, statistical, and algorithmic foundations of data science.''
Participants discussed potential opportunities for collaboration between the relevant communities and investigated workforce challenges and infrastructure development.
Ultimately, the meeting yielded a report
%~\cite{tfods_web_page,tfods_report} 
that suggested the creation of a ``center or institute, funded by an agency such as the NSF \ldots\; that emphasizes the foundational aspects of data~science.''\footnote{\url{http://www.cs.rpi.edu/TFoDS/TFoDS_v5.pdf}}
Much discussion during TFoDS addressed a \emph{justification challenge}, whereby foundations should inform and be informed by very practical problems (i.e., avoid representing pure theory divorced from practice), while not having to justify existence in terms of its immediate usefulness in particular domains.

\textbf{Cultural Challenge.}
A lot of TFoDS conversation also focused on what form the aforementioned center or interdisciplinary institute could take, or whether an institute is even the best mechanism to push the area forward.  However, there was little consensus.
Unresolved questions include the following.
%\begin{enumerate}
%\item
Should such an institute run short meetings or long programs?
%\item
What are attendance expectations?
Should the institute be an ``incubator'' for projects that cut across those three areas? 
%\item
Would those projects be funded by ``internal'' proposal solicitations?
Should they be virtual or physically located at one place?
%\item
How would one encourage long-term interactions?
Should projects attempt to encourage ``risky'' research, particularly among younger researchers?
%\item
How do these desires relate to the time scales for publication, recognition, and reward?
%\end{enumerate}
Far from being administrative issues, these questions (and the lack of straightforward answers) highlight a \emph{cultural challenge}: structuring interactions between 
different
areas---given their very different styles and modes of interaction, in addition to the conflicting power dynamics within and between them---is an extremely complicated task.

While perhaps not immediately obvious, many seemingly innocuous decisions can provide a strong selection bias toward/against certain areas, when it comes to interdisciplinary efforts.
The following actions all possess the potential for such selection bias: encouraging interdisciplinary interactions without understanding conflicting recognition and reward requirements; 
expecting publication immediately after one has vocalized an idea versus after he/she has clarified all of the details;
deciding on the final form of a complex interdiscipinary effort more quickly versus more deliberately;
requiring attendance throughout most or all of a long program; and
listing authors alphabetically, based on contribution, or according to some other rule.
Ignoring these issues---albeit often inadvertently---invariably undermines
interdisciplinary efforts.
Encouraging the research community to address these concerns in ways that draw strength from the diversity of researchers, and do not undermine their cultural sensibilities, continues to be a main~challenge.

%\vspace{-3mm}
\paragraph{Broader NSF Context.}

These discussions at the 2016 TFoDS Workshop occurred 
within a broader context of 
conversations at the NSF regarding the organization's long-term research agenda.
That same year, the NSF proposed its ``10 Big Ideas''\footnote{\url{https://www.nsf.gov/news/special_reports/big_ideas/}}, a set of 
``long-term research and process ideas that identify areas for future investment at the frontiers of science and engineering.'' 
One of these ideas---Harnessing the Data Revolution (HDR)\footnote{\url{https://www.nsf.gov/cise/harnessingdata/}}---focuses on both fundamental research in data science and engineering as well as the development of a 21st-century data-capable workforce, both to help researchers exploit the Big Data revolution.
This was part of the NSF's effort toward ``Growing Convergence Research,'' another big idea that seeks to integrate multiple disciplines to advance scientific discovery and innovation. 
The timing was right, and 
the NSF's first major investment toward HDR 
was the TRIPODS program.\footnote{\url{https://www.nsf.gov/news/news_summ.jsp?cntn_id=242888}}  %~\cite{nsf_www_tripods}. 

With TRIPODS, and following the suggestions of the TFoDS Workshop report, the NSF made a call for ``institutes'' on the foundations of data science.
However, the NSF was not overly-prescriptive as to what that means.
Instead, it split the program into two phase, the Phase~I and Phase~II (previously described) so that the community could determine what it desired from 
an institute that spanned three culturally quite different areas.
This two-phase structure also permits a ramp-up period before full-scale institute activities begin.
Possible institute-like activities include: 
research, education, workforce development, visitor hosting, and direction setting.
Phase~I PIs are addressing additional challenges, such as how to design institutes that do not grate against the standards of one or more of the communities and instead yield a true synergy of all of the three disciplines' best capabilities.

Each of the 12 Phase~I institutes approaches these challenges and its mission in somewhat different ways, acting as a type of ``experimental trial'' for struggles and successes. 
One of TRIPODS' more unusual aspects is the occurrence of a monthly PI call and annual PI meeting.
These allow for frank discussion of what does and does not work at different Phase~I institutes and in the general community.
Indeed, one of TRIPODS' most valuable aspects is its facilitation of a camaraderie between leading researchers with different backgrounds interested in similar challenges.

%\vspace{-3mm}
\paragraph{Interdisciplinary and Antedisciplinary Balance.}

TRIPODS' emphasis on interdisciplinary foundations and its addressing of cultural challenges associated with cross-cutting research are discussion points at the PI calls and meetings.

The three core areas of TRIPODS---statistics, mathematics, and theoretical computer science---address questions relevant to the theoretical foundations of data science, but they do so in very different and sometimes incomparable ways.
Each of the 12 Phase~I institutes is adopting its own approach. 
The FODA Institute at UC Berkeley will initially focus on four deep theoretical challenges: 
the possibility of a general complexity theory of inference in the context of optimization; 
the power of stability as a computational-inferential principle; 
the value of randomness as a statistical and algorithmic resource in data-driven computational mathematics; and 
the principled combination of science-based and data-driven models.
These foundational challenges straddle existing cultures of disciplinary research;
each is situated squarely at the interface of theoretical computer science, theoretical statistics, and applied mathematics; and
each is directly relevant to a wide range of very practical data science problems.

From the perspective of addressing cultural challenges associated with cross-cutting research, TRIPODS and the accompanying ``TRIPODS model'' offer the very real possibility of a ``test case'' on engineering funding for what Sean Eddy (Harvard University) calls ``antedisciplinary science'' (where ``ante'' means ``before,'' not ``anti'' as in ``against,'' although the two are clearly related).
He defines this as ``the science that precedes the organization of new disciplines, the Wild West frontier stage that comes before the law arrives''~\cite{eddy_antedisciplinary}.
The ways in which data science will evolve, e.g., whether the field will look more like present-day ``computational science'' or ``computer science,'' is still to be determined.
Nevertheless, Eddy's discussion about the National Institutes of Health's funding of computational biology is relevant to the discussion of cross-cutting research in the foundations of data and beyond:
``Focusing on interdisciplinary teams instead of interdisciplinary people reinforces standard disciplinary boundaries rather than breaking them down,'' he writes.  ``An interdisciplinary team is a committee in which members identify themselves as an expert in something else besides the actual scientific problem at hand, and abdicate responsibility for the majority of the work because it's not in their~field.''  

Many TRIPODS PIs are wrestling with these challenges, both in their own research agendas and their efforts to create Phase~II TRIPODS institutes that are broadly useful to the community.
Are the specifics of the current Phase~I -- Phase~II structure the best way to coalesce community understanding of what should comprise a longer-term institute (either for foundations of data or more general cross-cutting challenges)?
How can such institutes encourage interdisciplinary people as well as interdisciplinary teams?
How can they highlight the broad usefulness of interdisciplinary foundational work, without diluting its foundational content?
How can we ensure that current design decisions do not deter substantial participation by one of the disciplines of interest?
Of course, more obvious issues---ensuring that the whole is more than the sum of its parts and exploring novel ways to extend NSF funding---are also discussed.

While many challenging questions remain, participants of the current TRIPODS program are diligently wrestling with these questions to establish the foundations of data science.

%-----------------------------------------------------------------------
% %\newpage
% \bibliographystyle{unsrt}
% %\setcounter{page}{1}
% %\bibliography{PetrosRefs,communities,mwmbib_book,mwmbib_drft,mwmbib_jrnl,mwmbib_misc,mwmbib_proc,dnn,kbouchABI}
% %\bibliography{mwmbib_book,mwmbib_drft,mwmbib_jrnl,mwmbib_misc,mwmbib_proc}
% \bibliography{mwmbib_book,mwmbib_jrnl,mwmbib_misc}
%-----------------------------------------------------------------------

%-----------------------------------------------------------------------
\end{document}